%% file: main.tex
\documentclass[conference]{IEEEtran}
\usepackage[shortlabels]{enumitem}
\usepackage[nospace,noadjust]{cite}
\usepackage{graphicx} 
\usepackage{epsfig} 
\usepackage{amsmath,amsthm,amssymb}  
\usepackage{float}
\usepackage{cases,setspace,adjustbox}
\usepackage{array}
\usepackage[update,prepend]{epstopdf}
\usepackage{eqparbox}
\usepackage{url}
\usepackage{caption, subcaption}
\usepackage{xcolor}
\usepackage{mathtools}
\usepackage{bbm}
\usepackage{multirow}
\usepackage{multicol}

\usepackage[acronym]{glossaries}
\usepackage{comment}
\usepackage{xcolor}
\newcommand*\circled[1]{\kern-2.5em%
  \put(0,4){\color{white}\circle*{11}}\put(0,4){\circle{11}}%
  \put(-3,0){\color{black}\large#1}~~}
\input{glossary}
\newcommand{\ranA}{\text{LTE}}
\newcommand{\ranB}{\text{NR}}
\newcommand{\OPTP}{\text{OPT}_\text{max}}
\newcommand{\OPTA}{\text{OPT}_\text{avg}}
\begin{document}
\title{ProSAS: An O-RAN Approach \\to Spectrum Sharing between NR and LTE}
\author{Sneihil Gopal$^{\ddagger\dagger}$, David Griffith$^{\star}$, Richard A. Rouil$^{\star}$ and Chunmei Liu$^{\star}$\\
$^{\ddagger}$PREP Associate, National Institute of Standards and Technology (NIST), USA\\
$^{\dagger}$Department of Physics, Georgetown University, USA\\
$^{\star}$National Institute of Standards and Technology (NIST), USA\\
Emails: \{sneihil.gopal, david.griffith, richard.rouil, chunmei.liu\}@nist.gov}
\maketitle
\input{abstract}
\input{Introduction}
\input{Related_Work}
\input{Proposed_Framework}
\input{Results}
\input{Conclusion}
\begin{spacing}{0.925}
\bibliographystyle{IEEEtran}
\bibliography{references}
\end{spacing}
\appendix
\input{Appendix}
\end{document}

%% file: glossary.tex
\setacronymstyle{long-short}

\newacronym{AR}{AR}{Augmented Reality}
\newacronym{arima}{ARIMA}{Autoregressive Integrated Moving Average}

\newacronym{cdf}{CDF}{cumulative distribution function}
\newacronym{cnn}{CNN}{Convolutional Neural Network}
\newacronym{comp}{CoMP}{Coordinated Multi-Point}
\newacronym{convlstm}{ConvLSTM}{Convolutional LSTM}
\newacronym{cp}{CP}{Cyclic Prefix}
\newacronym{crs}{CRS}{Cell Specific
Reference Signal}
\newacronym{ctl}{CTL}{Communications Technology Laboratory}

\newacronym{d2d}{D2D}{Device-to-Device}
\newacronym{dc}{DC}{dual connectivity}
\newacronym{dci}{DCI}{Downlink Control Information}
\newacronym{dl}{DL}{downlink}
\newacronym{dsrc}{DSRC}{Dedicated short-range communications}

\newacronym{ets}{ETS}{Exponential Smoothing}
\newacronym{embb}{eMBB}{enhanced mobile broadband}
\newacronym{enb}{eNB}{Evolved Node-B}
\newacronym{epc}{EPC}{Evolved Packet Core}
\newacronym{emtc}{eMTC}{enhanced Machine-Type Communication}

\newacronym{fdd}{FDD}{Frequency Domain Duplex}
\newacronym{fdm}{FDM}{Frequency Domain Multiplexing}
\newacronym{firstnet}{FirstNet}{First Responder Network Authority}

\newacronym{gan}{GAN}{Generative Adversarial Network}
\newacronym{gps}{GPS}{Global Positioning System}
\newacronym{gnb}{gNB}{Next Generation Node-B}

\newacronym{harq}{HARQ}{Hybrid Automatic Repeat reQuest}

\newacronym{iiot}{IIoT}{Industrial Internet of Things}
\newacronym{its}{ITS}{Intelligent Transport Systems}
\newacronym{itu}{ITU}{International Telecom Union}

\newacronym{lte}{LTE}{Long Term Evolution}
\newacronym{lteapro}{LTE-A~Pro}{LTE Advanced Pro}
\newacronym{lstm}{LSTM}{Long Short-Term Memory}

\newacronym{mcs}{MCS}{Modulation and Coding Scheme}
\newacronym{mcptt}{MCPTT}{Mission-Critical Push-to-Talk}
\newacronym{mimo}{MIMO}{Multiple-Input Multiple-Output}
\newacronym{mmtc}{mMTC}{massive Machine-Type Communication}
\newacronym{mtc}{MTC}{Machine-Type Communication}
\newacronym{mlp}{MLP}{Multilayer Perceptron}
\newacronym{ma}{MA}{Moving Average}
\newacronym{mm}{MM}{Moving Median}
\newacronym{mae}{MAE}{Mean Absolute Error}

\newacronym{nist}{NIST}{National Institute of Standards and Technology}
\newacronym{nbiot}{NB-IoT}{Narrowband-Internet of Things}
\newacronym{nr}{NR}{New Radio}
\newacronym{ntia}{NTIA}{National Telecommunications and Information Administration}

\newacronym{oran}{O-RAN}{Open Radio Access Network}
\newacronym{owl}{OWL}{Online Watcher of LTE}

\newacronym{prosas}{ProSAS}{Proactive Spectrum Adaptation Scheme}
\newacronym{prb}{PRBs}{Physical Resource Blocks}
\newacronym{psc}{PSC}{Public Safety Communications}
\newacronym{pscch}{PSCCH}{Physical Sidelink Control Channel}
\newacronym{psfch}{PSFCH}{Physical Sidelink Feedback Channel}
\newacronym{pssch}{PSSCH}{Physical Sidelink Shared Channel}

\newacronym{qos}{QoS}{Quality of Service}

\newacronym{ran}{RAN}{Radio Access Network}
\newacronym{rat}{RAT}{Radio Access Technology}
\newacronym{re}{RE}{Resource Element}
\newacronym{ric}{RIC}{RAN Intelligent Controller}
\newacronym{relu}{ReLU}{Rectified Linear Unit}
\newacronym{rmse}{RMSE}{root-mean-square error}
\newacronym{rnn}{RNN}{Recurrent Neural Network}
\newacronym{rl}{RL}{Reinforcement Learning}

\newacronym{sdr}{SDR}{Software Defined Radio}
\newacronym{sl}{SL}{sidelink}
\newacronym{smo}{SMO}{Service and Management Orchestration}

\newacronym{tdd}{TDD}{Time Domain Duplex}
\newacronym{tdm}{TDM}{Time Domain Multiplexing}
\newacronym{timegan}{TimeGAN}{Time-Series Generative Adversarial Network}
\newacronym{tr}{TR}{Technical Report}

\newacronym{urllc}{URLLC}{ultra-reliable low-latency communication}
\newacronym{uav}{UAVs}{Unmanned Aerial Vehicles}
\newacronym{ue}{UE}{User Equipment}
\newacronym{ul}{UL}{uplink}

\newacronym{v2v}{V2V}{Vehicle-to-Vehicle}
\newacronym{v2x}{V2X}{Vehicle-to-Everything}
\newacronym{VR}{VR}{Vitual Reality}

\newacronym{wnd}{WND}{Wireless Networks Division}
\newacronym{wlan}{WLAN}{Wireless Local Area Network}
\newacronym{3gpp}{3GPP}{3rd Generation Partnership Project}

%% file: abstract.tex
\section*{Abstract}
\label{sec:abstract}
The Open Radio Access Network (O-RAN), an industry-driven initiative, utilizes intelligent Radio Access Network (RAN) controllers and open interfaces to facilitate efficient spectrum sharing between LTE and NR RANs. In this paper, we introduce the Proactive Spectrum Adaptation Scheme (ProSAS), a data-driven, O-RAN-compatible spectrum sharing solution. ProSAS is an intelligent radio resource demand prediction and management scheme for intent-driven spectrum management that minimizes surplus or deficit experienced by both RANs. We illustrate the effectiveness of this solution using real-world LTE resource usage data and synthetically generated NR data. Lastly, we discuss a high-level O-RAN-compatible architecture of the proposed solution.

%% file: Introduction.tex
\section{Introduction}
\label{sec:intro}
The surge in multimedia applications and the rapid digitization of many industries have led to increased mobile data usage in recent years~\cite{ericsson_2022}. To address this growing need for data connectivity, the \gls{3gpp} is deploying the fifth generation (5G) radio interface known as \gls{nr}~\cite{NR_Parkvall_2017}. \gls{nr} is expected to meet the rising demand for data traffic and enhance wireless connectivity for various emerging applications such as automotive, healthcare, public safety, and smart cities. It is engineered to deliver high data rates, low latency, and extensive coverage and will operate within two frequency bands: Frequency Range 1 (FR1), from 410~MHz to 7.125~GHz (also known as the sub-6 GHz band), and Frequency Range 2 (FR2), from 24.25~GHz to 52.6~GHz (also called the mmWave band). As most low-frequency bands are already allocated to 4G \gls{lte} networks, network operators face the challenge of either acquiring new spectrum or repurposing existing \gls{lte} spectrum for \gls{nr}. Both options entail substantial costs. Moreover, since \gls{lte} will continue to handle most data traffic in the near term, reallocating low-frequency bands from \gls{lte} to \gls{nr} without a concurrent expansion of \gls{nr}-capable devices will congest \gls{lte} bands and degrade network performance. Hence, operators need to strategically utilize their existing 4G infrastructure during the transition to \gls{nr} while providing services for legacy devices. This is necessary as \gls{lte} and \gls{nr} networks are expected to coexist for the foreseeable future. 

To ensure a smooth transition from \gls{lte} to \gls{nr} networks while supporting legacy devices and maintaining network performance, \gls{3gpp} has proposed a comprehensive set of solutions~\cite{ref:gopal2022}. These include \gls{lte}-compatible \gls{nr} numerology with a 15~kHz subcarrier spacing for unified time/frequency resource grids. Also, solutions include resource reservation, and \gls{dl} subcarrier puncturing to support \gls{emtc} (\gls{tr} 37.823), and mechanisms for resource allocation within NR carriers for \gls{nbiot} (\gls{tr} 37.824). 
Lastly, to help mitigate and manage interference for applications like V2X (\gls{tr} 37.985), \gls{3gpp} has introduced guard bands, dynamic and static power sharing mechanisms, and requirement-based access technology prioritization.

While these solutions lay a strong foundation for \gls{lte} and \gls{nr} network coexistence, real-world scenarios, diverse network environments, regulations, and evolving technological demands necessitate additional adaptations. The \gls{oran} initiative, with its open interfaces, virtualization, and intelligence, supports innovation within networks. Equipped with Artificial Intelligence/Machine Learning (AI/ML)-capable \gls{ric}, \gls{oran} enables advanced spectrum sharing. Within this framework, we propose \gls{prosas}, a data-driven, O-RAN-compatible resource-sharing solution. \gls{prosas} capitalizes on \gls{oran}'s capabilities, focusing on intelligent demand prediction and resource allocation to minimize resource surpluses or deficits experienced by networks. 

We utilize statistical models and deep learning techniques to analyze and predict radio resource demand patterns. This analysis uses real-world \gls{lte} resource usage data collected using BladeRF \gls{sdr}\footnote{Certain commercial equipment, instruments, or materials are identified in this paper in order to specify the experimental procedure adequately. Such identification is not intended to imply recommendation or endorsement by National Institute of Standards and Technology (NIST), nor is it intended to imply that the materials or equipment
identified are necessarily the best available for the purpose.\label{refnote}} and \gls{owl}, an open-source \gls{lte} sniffer~\cite{ref:owl_bui_2016}, as well as \gls{nr} data generated synthetically using \gls{timegan}~\cite{ref:timegan}. The proposed scheme is \gls{oran}-compatible and can be implemented within the \gls{oran} architecture as follows: the modeling and training processes corresponding to demand prediction can occur within the non-real-time (non-RT) \gls{ric}, leveraging data gathered from \gls{ran}. The optimal model can be selected and subsequently deployed to the near-real-time (near-RT) \gls{ric} for inference. During the inference phase, the near-RT \gls{ric} can allocate resource blocks to minimize any surplus or deficit for both \gls{ran}s. To demonstrate \gls{prosas}'s compatibility with \gls{oran} requirements, we discuss a high-level architectural overview of the modules and the deployment scenario alongwith a summary of the end-to-end workflow involved. To the best of our knowledge, this paper is the first attempt to propose an \gls{oran}-compatible scheme that manages radio resources proactively and intelligently to balance the demands of \gls{lte} and \gls{nr} networks. Our use of real-world data and alignment with industry standards makes this proposal practical and pertinent for network operators. Furthermore, we provide a valuable set of \gls{lte} resource usage data for researchers, and we demonstrate the utility of a GAN in generating synthetic radio resource usage data~\cite{ref:datset}. 

The structure of the paper is as follows: Section~\ref{sec:prior_work} delves into related work. Section~\ref{sec:proposed_framework} provides an overview of \gls{oran}, the \gls{prosas} framework, and the high-level architecture of the \gls{oran}-compatible deployment scenario. The effectiveness of \gls{prosas} is demonstrated in Section~\ref{sec:results}. Finally, Section~\ref{sec:conclusion} concludes the paper.

%% file: Related_Work.tex
\section{Related Work}
\label{sec:prior_work}
In addition to \gls{3gpp}'s work, the research community has also studied \gls{lte}-\gls{nr} coexistence~
\cite{ref:Coex_Xu_2021,ref:Coex_Alexandre_2019,ref:Coex_wan_2018,ref:Coex_Levanen_2018,ref:Coex_An_2020,ref:Coex_li_2021,ref:Coex_Wan_2019}. In~\cite{ref:Coex_Xu_2021}, the authors addressed co-channel interference, and proposed a low-complexity interference mitigation method for \gls{nr} devices. In~\cite{ref:Coex_Alexandre_2019}, the authors presented experimental results for \gls{lte}-\gls{nr} coexistence in the 700 MHz band, demonstrating peaceful \gls{dl} coexistence. In~\cite{ref:Coex_wan_2018}, the authors introduced a spectrum exploitation mechanism balancing transmission efficiency, coverage, and latency. In~\cite{ref:Coex_Levanen_2018}, the authors discussed \gls{lte}-\gls{nr} \gls{ul} coexistence, revealing that \gls{lte} needs a single resource block as a guard band, while \gls{nr} remains unaffected by \gls{lte}. In~\cite{ref:Coex_An_2020}, the authors explored \gls{lte} \gls{fdd} and \gls{nr} \gls{fdd} coexistence in the 2.1~GHz band and showed that \gls{lte} \gls{fdd} \gls{dl} causes harmful interference to \gls{nr} \gls{fdd} \gls{dl}. Conversely,~\cite{ref:Coex_li_2021} analyzed \gls{nr} \gls{fdd} and \gls{lte} \gls{tdd} coexistence in the 1.8 GHz band, showing their compatibility in adjacent bands.~\cite{ref:Coex_Wan_2019} explored \gls{nr} \gls{ul}/\gls{dl} coexistence with \gls{lte} \gls{ul}/\gls{dl}, highlighting the role of \gls{ul} sharing in balancing spectrum efficiency, latency, coverage, and channel bandwidth. Additionally, \gls{lte}-\gls{nr} coexistence has been examined for specific applications like \gls{emtc}~\cite{ref:Coex_Ratasuk_2020}, \gls{nbiot}~\cite{ref:Coex_Mozaffari_2019}, and V2X~\cite{ref:V2X_Garcia_2021}. 

While the existing work on \gls{lte}-\gls{nr} coexistence has made significant strides, several aspects can be improved with the integration of \gls{oran} principles. One missing element is adaptability and intelligent management of resources to proactively balance the demands of \gls{lte} and \gls{nr} networks, which is our focus. We will show how this element can support spectrum sharing, which has recently gained the attention of the \gls{oran} Alliance Working Group 1 (WG1)~\cite{ref:oran_tr_wg1} and has been explored in several studies~\cite{ref:oran_smith_2021,ref:oran_mungari_2021_rl,ref:oran_baldesi_2022_charm,ref:oran_kulacz_2022_dynamic}. In~\cite{ref:oran_smith_2021}, the authors propose spectrum sharing between government Low Earth Orbit satellites and 5G \gls{ul} using \gls{oran} intelligence and open interfaces. In~\cite{ref:oran_mungari_2021_rl}, a \gls{rl}-based radio resource management solution is developed, leveraging the O-RAN platform.~\cite{ref:oran_baldesi_2022_charm} introduced \gls{oran}-compliant frameworks for enabling spectrum sharing. Specifically,~\cite{ref:oran_baldesi_2022_charm} demonstrated the framework's performance in \gls{lte}-WiFi coexistence in unlicensed bands, using srsRAN for prototype development and the Colosseum channel emulator for data collection.~\cite{ref:oran_kulacz_2022_dynamic} explored contextual user density data for spectrum management in O-RAN-based networks, offering numerical validation of the proposed concept. While these works make valuable contributions, \gls{prosas} features a data-driven, intelligent, and intent-driven approach to spectrum sharing and resource management. Its use of real-world data and compatibility with O-RAN makes it a potential tool for supporting \gls{lte}-\gls{nr} coexistence.

%% file: Proposed_Framework.tex
\section{The ProSAS Framework}
\label{sec:proposed_framework}
\gls{oran} represents a transformative approach to modernizing the traditional \gls{ran}. Unlike traditional \gls{ran}s, which often rely on proprietary, monolithic hardware and software solutions, \gls{oran} promotes open interfaces and virtualization technologies and introduces near-RT \gls{ric}s and non-RT \gls{ric}s to coordinate and manage network resources. These \gls{ric}s play pivotal roles in the \gls{oran} architecture by optimizing radio resource allocation, enhancing network performance, and ensuring quality of service (QoS). Furthermore, \gls{oran} promotes innovation by facilitating the development of \gls{ran} applications (rApps) and external applications (xApps) that can seamlessly interface with the \gls{ric}s and the \gls{ran} infrastructure, providing advanced functionalities, intelligence, and adaptability. We have designed \gls{prosas} such that it can use rApps, xApps, and open interfaces, and therefore harness the capabilities of the \gls{oran} architecture.

In this paper, we propose an intelligent radio resource demand prediction and management scheme for intent-driven (priority-based) spectrum management. We employ a high-level model for \gls{lte} and \gls{nr} that makes few assumptions about the underlying technology. We assume that \gls{nr} employs an \gls{lte}-compatible numerology with 15~kHz subcarrier spacing, resulting in identical time/frequency resource grids for both networks and we assume that both networks use a shared pool of time-frequency resources, referred to as \gls{prb}. Note that we define resource demand as the number of \gls{prb} required by a network for data transmission. Our approach learns the resource usage patterns of each network by analyzing their individual resource usage time-series datasets, using a variety of statistical models and deep learning architectures to predict future demand. We choose the method with the lowest \gls{rmse} as the most suitable predictor. Based on the predicted demand from the selected prediction model, we allocate resources to minimize any surplus or deficit experienced by both networks, where a surplus indicates resource over-provisioning and a deficit represents under-provisioning, leading to packet loss or buffering. Next, we discuss the proposed framework and its high-level integration into the \gls{oran} architecture.

\subsubsection{Radio Resource Demand Prediction}
Understanding radio resource usage patterns is crucial for effective resource management because network operators can then mitigate the costs of unexpected spikes in data traffic. We use a toolkit of time-series predictive models that incorporates statistical methods and deep learning architectures. It includes the \gls{arima}~\cite{ref:time_series_book_nielsen_2019} model, which combines autoregressive (AR) and moving average (MA) components with differencing to handle linear dependencies in the time series data. We also use the toolkit's \gls{ets} model~\cite{ref:time_series_book_nielsen_2019} to capture data patterns, including trends and seasonality.

In addition to statistical models, our toolkit includes three deep learning models tailored for time series forecasting. The \gls{mlp} Model~\cite{ref:time_series_book_brownlee_2018_deep} is a neural network with interconnected layers of neurons, well-suited for uncovering intricate data patterns in sequential data. The \gls{cnn} Model~\cite{ref:time_series_book_brownlee_2018_deep}, originally designed for image data, can also be applied to time series data, using convolutional layers to learn relevant features, and is useful for tasks emphasizing local patterns. Lastly, the \gls{lstm} Model~\cite{ref:time_series_book_brownlee_2018_deep}, a category of \gls{rnn}, engineered to capture long-term dependencies in sequential data. Its ability to retain information over extended time intervals makes it effective for time series forecasting tasks. Our toolkit also encompasses variations of the \gls{lstm} model~\cite{ref:time_series_book_brownlee_2018_deep}, namely, \gls{cnn}-\gls{lstm} (combining the convolutional layers of \gls{cnn} with the sequential modeling capability of \gls{lstm}), and \gls{convlstm} (integrating convolutional operations within \gls{lstm} units). 

To establish a performance baseline for these models, we incorporate simpler models, including the Persistence or Naive model, \gls{ma} model, and \gls{mm} model. The Persistence model sets the forecast for the next time step as the previous time step, providing a benchmark for more complex models. The \gls{ma} model calculates the forecast as the average of the previous $n$ observations, which captures short-term trends or fluctuations. Similarly, the \gls{mm} model calculates the forecast as the median of the previous $n$ observations, offering robustness against outliers and generating a smoother forecast. This set of models equips us to tackle diverse aspects of radio resource usage pattern analysis, capturing simple trends and complex dependencies in time series data. 

While several metrics, such as computational time and \gls{mae}, are used to assess prediction model performance, we rely on the widely adopted \gls{rmse} metric \cite{ref:time_series_book_brownlee_2018_deep}. \gls{rmse} quantifies the standard deviation of prediction errors and is defined as: $\text{\gls{rmse}} = \bigl[(1/n) \sum\nolimits_{i=1}^{n} (|(f_{i} - \hat{f_{i}}|)^2\bigr]^{1/2}$, where $n$ is the number of time steps in a data set and $f_{i}$ and $\hat{f_{i}}$ are respectively the observed and predicted resource usage at the $i$th time step. 

\subsubsection{Optimal Resource Allocation}
We now describe how resources from a shared pool of \gls{prb} are distributed between coexisting \gls{lte} and \gls{nr} networks. We assume that the controller allocates \gls{prb} to meet each network's resource demands at regular intervals. These intervals can represent time slots in an \gls{lte} system, $1$~ms subframes, periods of several seconds, minutes, or hours, depending on the scenario. 

Let $N_{R}$ represent the size of the \gls{prb} resource pool used by the controller to distribute resources ($N_A$ for $\ranA$ and $N_B$ for $\ranB$). The primary goal is to minimize resource surplus or deficit for both networks. Our scheme determines the optimal allocation by evaluating a performance metric that addresses the networks' varying demands. Although the scheme can support static allocation (performed once during network setup) or dynamic allocation (repeated at granularities as small as 1 ms), we focus on quasi-static resource allocation. Since resource allocation follows demand prediction, we assume that the controller has knowledge of the networks' future resource demands for a defined period (e.g., the next hour or minute) and allocates resources accordingly. 

Next, we define the optimization problem (OPT) as follows:
\begin{align}
\textbf{OPT:}&\min_{N_{A}, N_{B}} J(N_{A}, N_{B} | \gamma, x_{D_{A}}, x_{D_{B}}),\nonumber\\
\textbf{s.t.}&\quad N_{A} + N_{B} \leq N_{R},\nonumber\\
&\quad 0 \leq N_{A} \leq N_{R}, \nonumber\\
&\quad 0 \leq N_{B} \leq N_{R},
\label{eqn:opt_prob}   
\end{align}
where, 
\begin{equation}
J= \gamma\left(\frac{N_A-x_{D_A}}{x_{D_A}}\right)^2 
+ (1-\gamma) \left(\frac{N_B-x_{D_B}}{x_{D_B}}\right)^2,
    \label{eqn:obj_func_gen}
\end{equation}
is the weighted sum of squared fractional surpluses/deficits, i.e., $(N_A-x_A)/x_A$ and $(N_B-x_B)/x_B$, experienced by $\ranA$ and $\ranB$, respectively. $(N_A, N_B)$ is the ordered pair of allocation and $(x_{D_{A}},x_{D_{B}})$ is the ordered pair containing statistics associated with the distributions of $\ranA$'s and $\ranB$'s respective demands, i.e., ($D_{A,i}, D_{B,i}$), for $i\in[1,n]$, where $i$ is the time in a length-$n$ period. $\gamma \in (0,1)$ is the weighting factor that enables intent-driven spectrum management by allowing the controller to prioritize one network over the other. Specifically, $\gamma \rightarrow 1$ assigns a higher priority to $\ranA$, whereas, $\gamma \rightarrow 0$ assigns a higher priority to $\ranB$. 

We consider two variations of OPT: (a) $\OPTP$ which uses the maximum network demand ($M_{D_A}, M_{D_B}$), as the network statistics ($x_{D_A}, x_{D_B}$) in Eq.~\eqref{eqn:obj_func_gen}, further substituted in Eq.~\eqref{eqn:opt_prob}, and (b) $\OPTA$ which uses the mean of the objective function, i.e., $E[J]$, and therefore considers the mean and variance of each network's demand. 

For $\OPTA$, the objective function is:
\begin{align}
J_{avg} &= E\left[\gamma\left(\frac{N_A-D_A}{D_A}\right)^2 
+ (1-\gamma) \left(\frac{N_B-D_B}{D_B}\right)^2\right], \nonumber\\
&= 1+ \gamma N_{A}\,C_A + (1-\gamma) N_{B}\, C_B,
\label{eq:obj_func_avg}
\end{align}%
where, by linearity of the expectation $E[\cdot]$, for $X \in \{A,B\}$,

\begin{equation}
C_X = N_{X} \left(\frac{1}{\mu_{D_{X}}^2} + \frac{3\sigma_{D_{X}}^2}{\mu_{D_{X}}^4} \right) -2\left(\frac{1}{\mu_{D_{X}}} + \frac{\sigma_{D_{X}}^2}{\mu_{D_{X}}^3} \right), 
\end{equation}
$\mu_{D_{A}}$ and $\mu_{D_{B}}$ are the mean, and $\sigma_{D_{A}}^2$ and $\sigma_{D_{B}}^2$ are the variance of $\ranA$ and $\ranB$'s demand, respectively. As with $\OPTP$, $\OPTA$ can be obtained by substituting Eq.~\eqref{eq:obj_func_avg} in Eq.~\eqref{eqn:opt_prob}. $\OPTP$ and $\OPTA$ are both convex-optimization problems and each has a global minimum.

For the optimal resource pool partition $(N_A^*,N_B^*)$, we compute the average normalized surplus or deficit, denoted by $S_{A}$ and $S_{B}$ for $\ranA$ and $\ranB$, respectively:
\begin{equation}
\small
S_{A} = \frac{1}{n}\sum\limits_{i=1}^{n}\left(\frac{N_A^{*}-D_{A,i}}{D_{A,i}}\right),\enspace
S_{B} = \frac{1}{n}\sum\limits_{i=1}^{n}\left(\frac{N_B^{*}-D_{B,i}}{D_{B,i}}\right)\,.
\label{eq:surplus_deficit}
\normalsize
\end{equation}
Lastly, we compute the fairness of the allocations for $\OPTP$ and $\OPTA$ using Jain's Fairness Index defined as:
\begin{equation}
\small
    F = \frac{(N_A^*+N_B^*)^2}{2\bigl((N_A^*)^2+(N_B^*)^2\bigr)} .
    \label{eq:JFI}
\normalsize
\end{equation}
\subsubsection{Deployment Architecture}
\begin{figure}[t]
    \centering
    \includegraphics[width=\columnwidth]{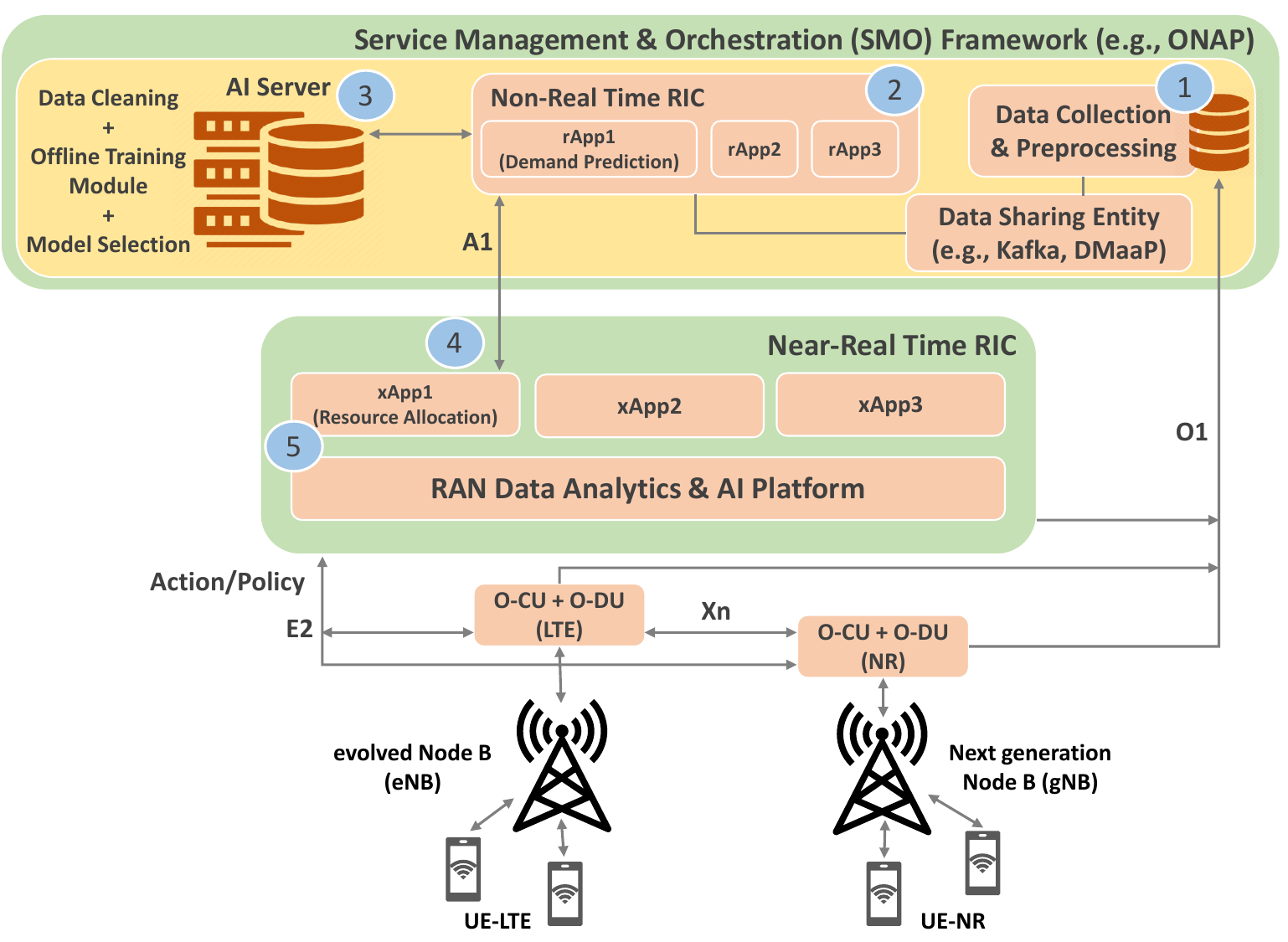}
    \vspace{-0.5em}
    \caption{\small High-level structure illustrating the deployment of ProSAS within the \gls{oran} architecture~\cite{ref:oran_tr_wg1}.}
    \label{fig:deployment}
\vspace{-1.75em}
\end{figure}
Next, we describe how \gls{prosas} could be implemented in the \gls{oran} architecture, with the structure and end-to-end flow illustrated in Fig.~\ref{fig:deployment}. The design and deployment scenarios are based on the guidelines outlined in the TR~\cite{ref:oran_tr_wg2} by the \gls{oran} Alliance.
\begin{enumerate}[label=\protect\circled{\arabic*}]
\item Data collected from the \gls{ran} components (O-CU and O-DU), including the \gls{dci}, is transmitted to the data collector in the \gls{smo} entity through the O1 interface. This data is input for the Demand Prediction rApp within the non-RT \gls{ric}.
\item Using a data-sharing entity such as DMaaP or Kafka, the gathered data is routed to the non-RT \gls{ric} in the \gls{smo}.
\item The relevant AI/ML models, including statistical models and deep learning architectures, in the AI server (e.g., AcumosAI) in the \gls{smo}, are queried by the non-RT \gls{ric}. Once these models have been trained on the AI server, the non-RT \gls{ric} is notified of the inference.
\item The results of the inference and associated policies are forwarded to the Resource Allocation xApp in the near-RT \gls{ric} via the A1 interface.
\item The Resource Allocation xApp configures the resource allocation solution, which is derived by solving $\OPTP$ and $\OPTA$, and selected based on  Jain's fairness index. This solution is conveyed to the \gls{ran} components using the E2 interface. 
\end{enumerate}

%% file: Results.tex
\section{Numerical Results for ProSAS}
\label{sec:results}
\begin{figure}
     \centering
     \begin{subfigure}[b]{0.925\columnwidth}
         \centering
         \includegraphics[width=\columnwidth]{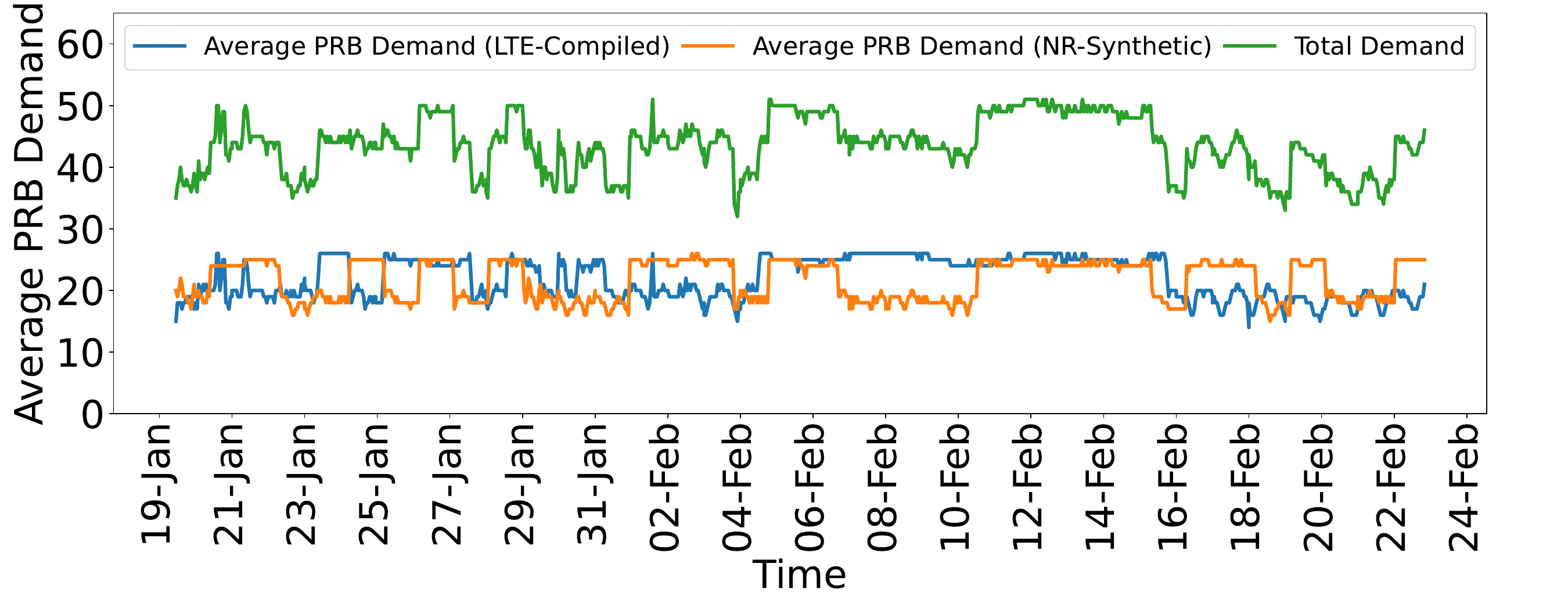}
     \end{subfigure}
     \begin{subfigure}[b]{0.925\columnwidth}
          \centering
         \includegraphics[width=\columnwidth]{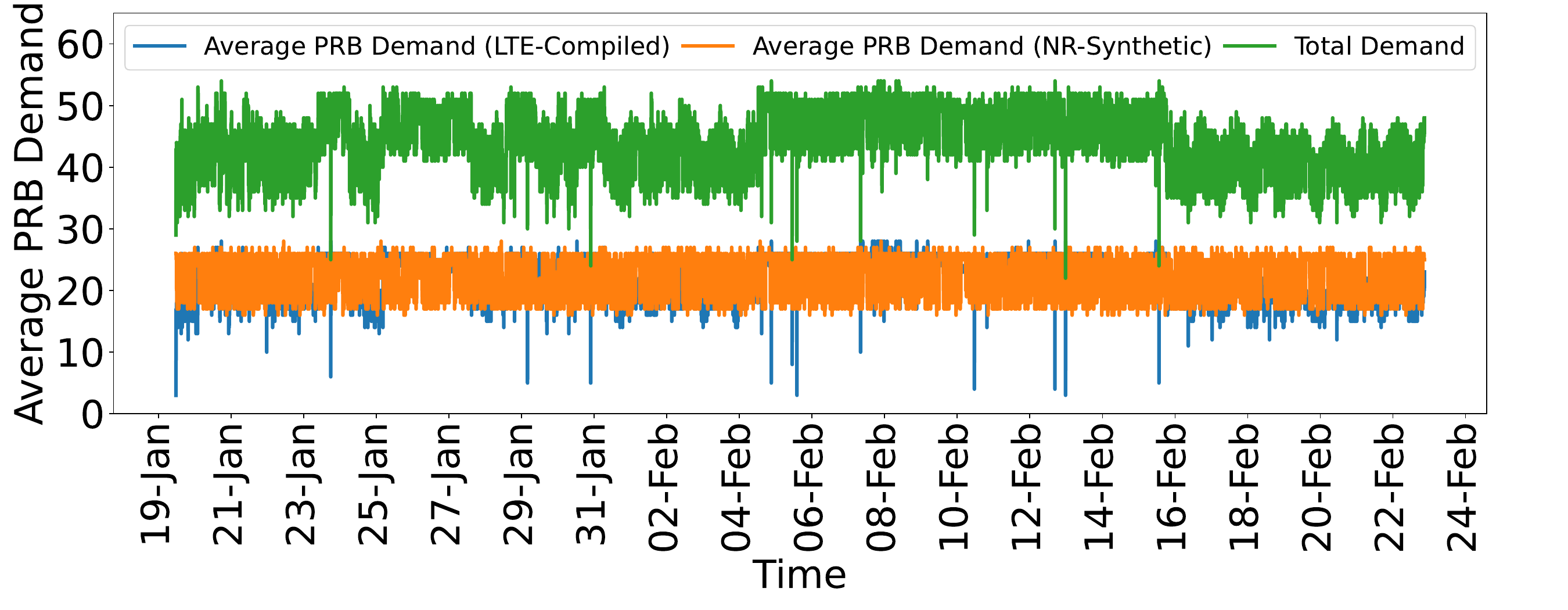}
     \end{subfigure}
     \vspace{-1em}
     \caption{\small Average PRB usage vs. time for \gls{lte}, \gls{nr} and \gls{lte}+\gls{nr}, i.e., aggregate, corresponding to time granularity of 1-hour and 1-minute.}
     \label{fig:lte_nr_demand}
\vspace{-1.55em}
\end{figure}
For our study, we compiled an extensive dataset of \gls{lte} scheduling information, collected at the NIST Gaithersburg campus between January and February 2023~\cite{ref:datset}. This dataset is from downlink traffic at 2115~MHz (Band~4) and was collected using OWL~\cite{ref:owl_bui_2016}, an online decoder of the \gls{lte} control channel. We employed a BladeRF SDR board to transmit the collected real-world \gls{lte} signals to a laptop running Ubuntu~20.04 and executing OWL to decode the signals. OWL captured the \gls{lte} \gls{dci} broadcast by the base station at the 1~ms subframe level. The dataset includes information such as the system frame number (SFN), subframe index, radio network temporary identifiers (RNTIs), the number of \gls{prb} allocated to devices, modulation and coding scheme (MCS), and DCI message type. We extracted the number of \gls{prb} corresponding to data transmissions, specifically \gls{dci} format~2B. We also included timestamps reflecting when this data was collected, resulting in a time series dataset detailing PRB allocations for the \gls{lte} network. Importantly, this data contains no user-specific information, ensuring complete anonymity.

Due to the lack of availability of tools that can capture \gls{nr} resource usage data, using the compiled real-world \gls{lte} dataset, we generated a synthetic PRB allocation dataset corresponding to the \gls{nr} network. To create this synthetic \gls{nr} dataset, we leveraged \gls{timegan}~\cite{ref:timegan}, a tool that generates synthetic time-series data by capturing temporal dependencies through \gls{rnn}. We used \gls{timegan} and the collected real-world LTE time-series data to produce synthetic time-series \gls{nr} data that closely mirrors the statistical properties and patterns observed in the \gls{lte} data. This approach is valid due to our assumption that \gls{nr} employs an \gls{lte}-compatible numerology with 15~KHz subcarrier spacing, and shares an identical time-frequency resource grid. We also assumed that \gls{nr} systems would serve \gls{lte} devices. 

For our analysis, we performed a resampling of the time-series data, transitioning from PRB allocation per millisecond to PRB allocation per hour and per minute by using the mean function. Consequently, we obtained datasets representing the average PRB allocation for both \gls{lte} and \gls{nr} per hour and minute. The former dataset comprises approximately 860~samples, while the latter encompasses approximately 49\,554~samples. We defer the analysis for time granularities less than 1~minute for future research. Fig.~\ref{fig:lte_nr_demand} displays the average PRB usage/demand over time for \gls{lte}, \gls{nr}, and the combined (\gls{lte}+\gls{nr}) datasets, both at 1-hour and 1-minute granularities. 
\begin{figure}
     \centering
     \begin{subfigure}[b]{0.48\columnwidth}
         \centering
         \includegraphics[width=0.98\columnwidth]{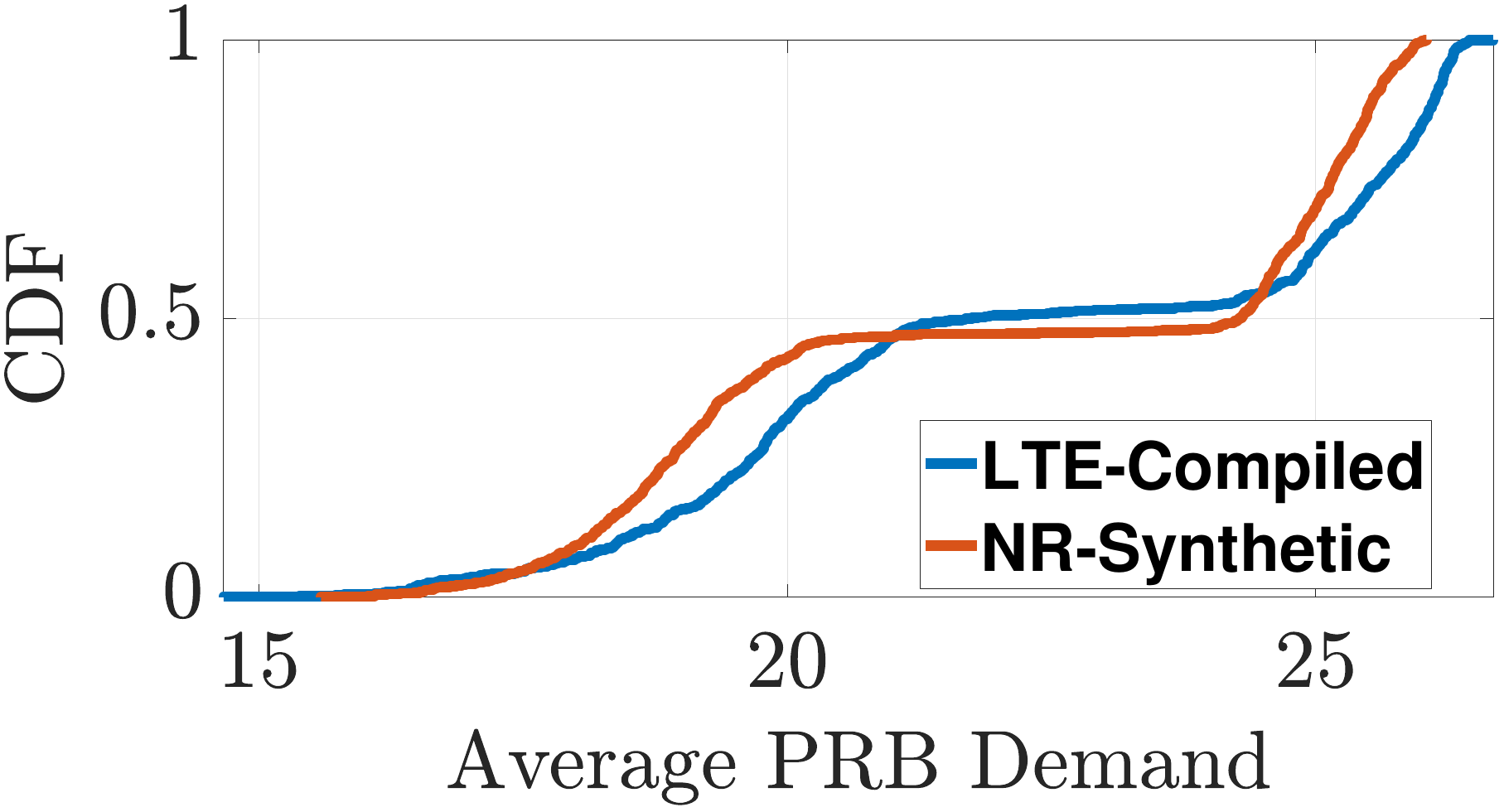}
         \caption{Granularity = 1 hour}
     \end{subfigure}\hfill
     \begin{subfigure}[b]{0.48\columnwidth}
          \centering
         \includegraphics[width=0.98\columnwidth]{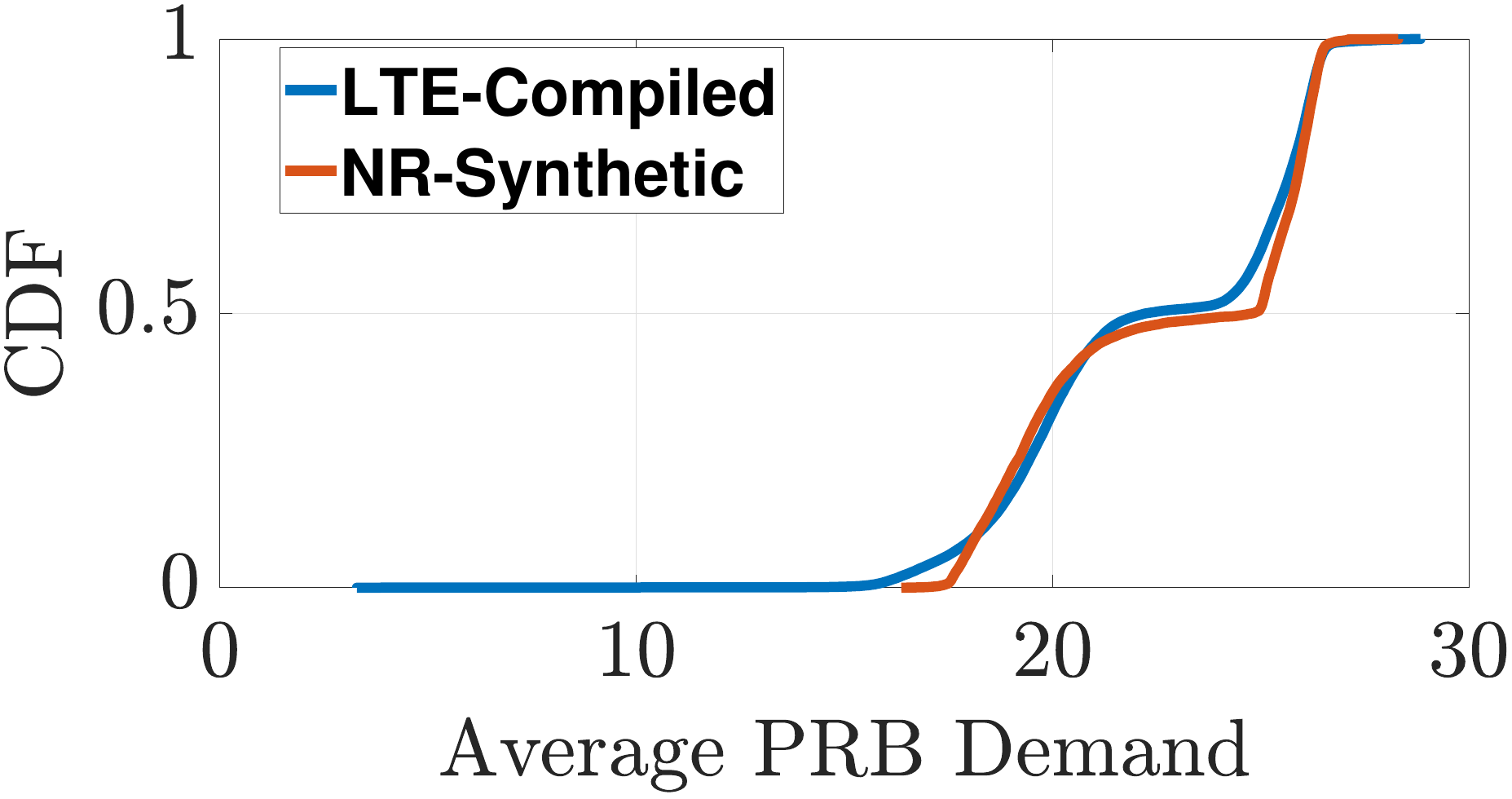}
         \caption{Granularity = 1 minute}
     \end{subfigure}
     \vspace{-0.5em}
     \caption{\small CDF of average PRB demand for \gls{lte} and synthetically generated \gls{nr} dataset for time granularity 1-hour and 1-minute.}
     \label{fig:cdf_lte_nr_demand}
\vspace{-1.6em}
\end{figure}
Additionally, Fig.~\ref{fig:cdf_lte_nr_demand} shows the \gls{cdf} of PRB demands from both networks at different granularities. This graph highlights the similarity in demand distribution between the compiled \gls{lte} dataset and the synthetically generated \gls{nr} dataset, thereby confirming the effectiveness of \gls{timegan}. Next, we delve into the outcomes of the various resource demand prediction models, in addition to those pertaining to the resource allocation schemes.
\begin{table*}[]
\footnotesize
    \centering
    \scalebox{0.875}{
    \begin{tabular}{|p{3.75em}|p{20em}|p{8.25em}|p{8.25em}|p{8.25em}|p{8.25em}|}
    \hline
        \textbf{Model} & \textbf{Training Hyperparameters} & \textbf{LTE (1-hour)} & \textbf{NR (1-hour)} &\textbf{ LTE (1-minute)} & \textbf{NR (1-minute)} \\
        \hline\hline
        Naive& (length of historical data, offset) & (1, 1) & (1, 1) & (1, 1) & (1, 1)\\
        MA  & (length of historical data, offset) & (2, 1) & (2, 1) & (2, 1) & (2, 1)\\
        MM & (length of historical data, offset) & (2, 1)  & (2, 1) & (2, 1) & (2, 1)\\
        \gls{arima} & ((no. of lag observations (p), degree of differencing (d), order of \gls{ma} (q)), trend)& ((1, 3, 2), `c') & ((3, 0, 3), `c') & ((0, 1, 3), `c')  & ((2, 1, 0), `c')\\
         ETS & (trend, damped trend, seasonal) & (`mul', False, None)& ('add', True, None) & (None, False, None)  & ('add', True, None) \\
         MLP & (inputs, nodes, epochs, batch size) & (1, 150, 100, 150) & (1, 150, 500, 100) & (10, 100, 100, 150)  & (1, 150, 500, 100) \\
         CNN & (inputs, filters, kernels, epochs, batch size) &(2, 256, 1, 100, 150) & (2, 256, 1, 500, 150) & (2, 256, 1, 500, 150) & (2, 256, 1, 500, 150)\\
         LSTM & (inputs, nodes, epochs, batch size) & (1, 150, 100, 100) & (1, 150, 500, 150) & (1, 100, 500, 100) & (1, 150, 500, 150) \\
         \gls{lstm}-I & (sequences, steps, filters, kernels, nodes, epochs, batch size) & (1, 2, 256, 1, 100, 100, 150) & (1, 2, 256, 1, 100, 100, 150) & (1, 2, 256, 1, 100, 500, 150)  & (1, 2, 256, 1, 100, 100, 150)\\
         \gls{lstm}-II & (sequences, steps, filters, kernels, nodes, epochs, batch size) & (2, 1, 256, 1, 100, 500, 150) & (2, 1, 256, 1, 100, 500, 150) & (2, 1, 256, 1, 150, 500, 100) & (2, 1, 256, 1, 100, 500, 150)\\
    \hline
    \end{tabular}}
    \caption{\small List of training hyperparameters chosen for each prediction model using the grid search framework~\cite{ref:time_series_book_brownlee_2018_deep}.}
    \label{tab:model_parameters}
\normalsize
\vspace{-1.5em}
\end{table*}

\begin{figure}
    \centering
    \includegraphics[width = 0.98\columnwidth]{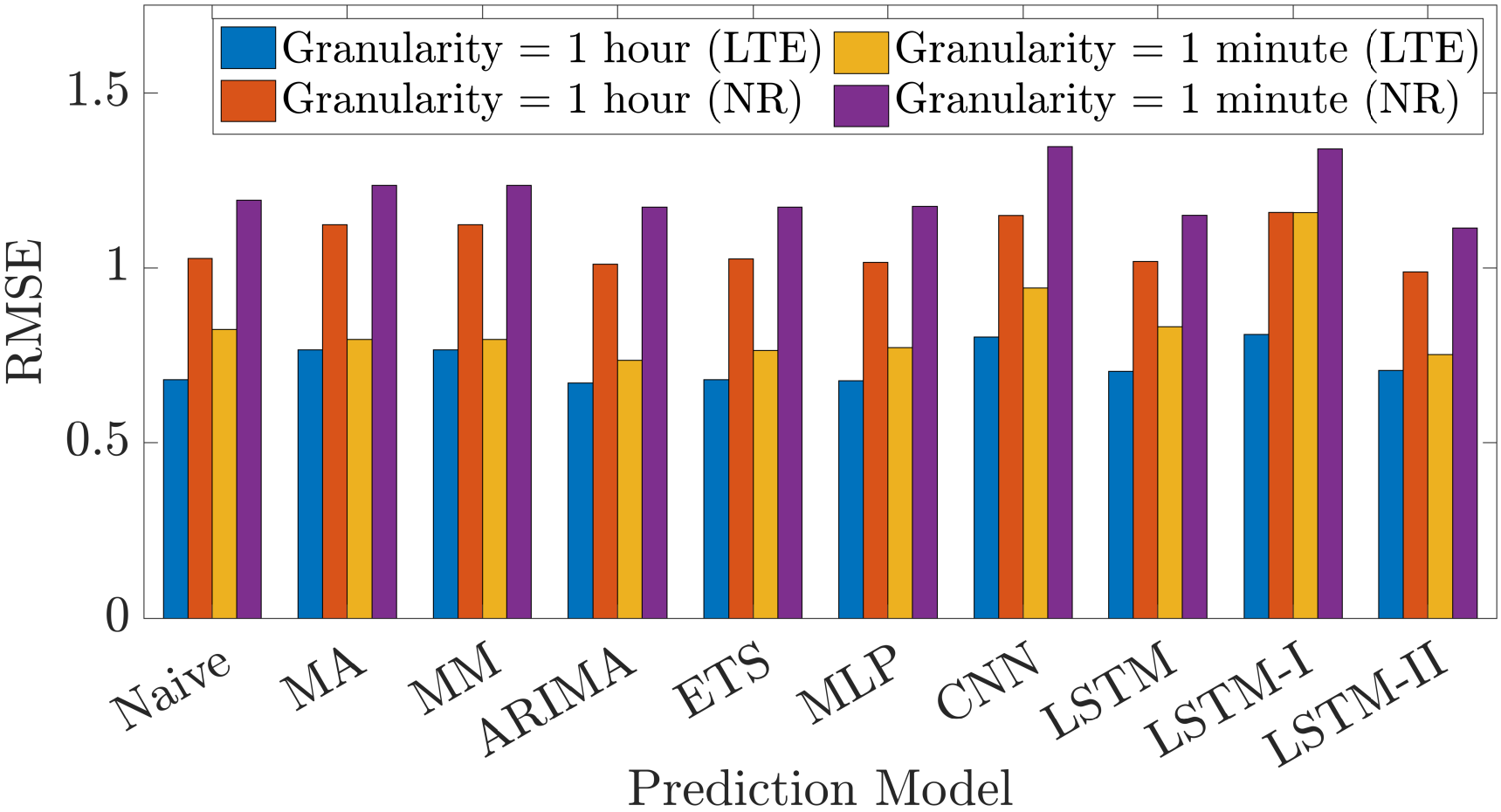}
    \vspace{-0.5em}
    \caption{\small Evaluation of prediction models in terms of \gls{rmse}. \gls{lstm}-I and \gls{lstm}-II refer to \gls{cnn}-\gls{lstm} and \gls{convlstm}, respectively.}
    \label{fig:rmse_comparison}
\vspace{-1.5em}
\end{figure}

As mentioned earlier, our comprehensive demand prediction toolkit comprises various models, combining both statistical approaches 
and deep learning architectures. 
We perform a comparative analysis, evaluating the performance of these models against alternatives like the Persistence or Naive model, \gls{ma} model, and \gls{mm} model. 
The implementation of these prediction models is carried out in Python, with Keras serving as the backend for deep learning architectures. For each model, we have designed a framework for grid-searching model hyperparameters, employing \gls{rmse} as our guiding metric through one-step walk-forward validation. The selected hyperparameters for each model are listed in Table~\ref{tab:model_parameters} for transparency and reproducibility. For training and validating the model architecture,  we use 66~\% for training at each granularity, i.e., 1-hour and 1-minute~\cite{ref:time_series_book_nielsen_2019}. Lastly, for the deep learning architectures, we use the Adam optimization algorithm~\cite{ref:time_series_book_nielsen_2019} and employ \gls{relu} as the activation function. In Fig.~\ref{fig:rmse_comparison}, we compare the prediction models based on \gls{rmse} for both networks and granularities. Notably, \gls{arima} outperforms other models for \gls{lte} data, whereas the \gls{convlstm} model has the lowest RMSE for \gls{nr} data. This highlights the need for tailored models rather than a one-size-fits-all approach. Also, our results indicate that while complex deep-learning architectures can be effective, simpler statistical models can also yield strong performance.

\begin{figure*}[t]
     \centering
     \begin{subfigure}[b]{0.31\textwidth}
         \centering
         \includegraphics[width=\textwidth]{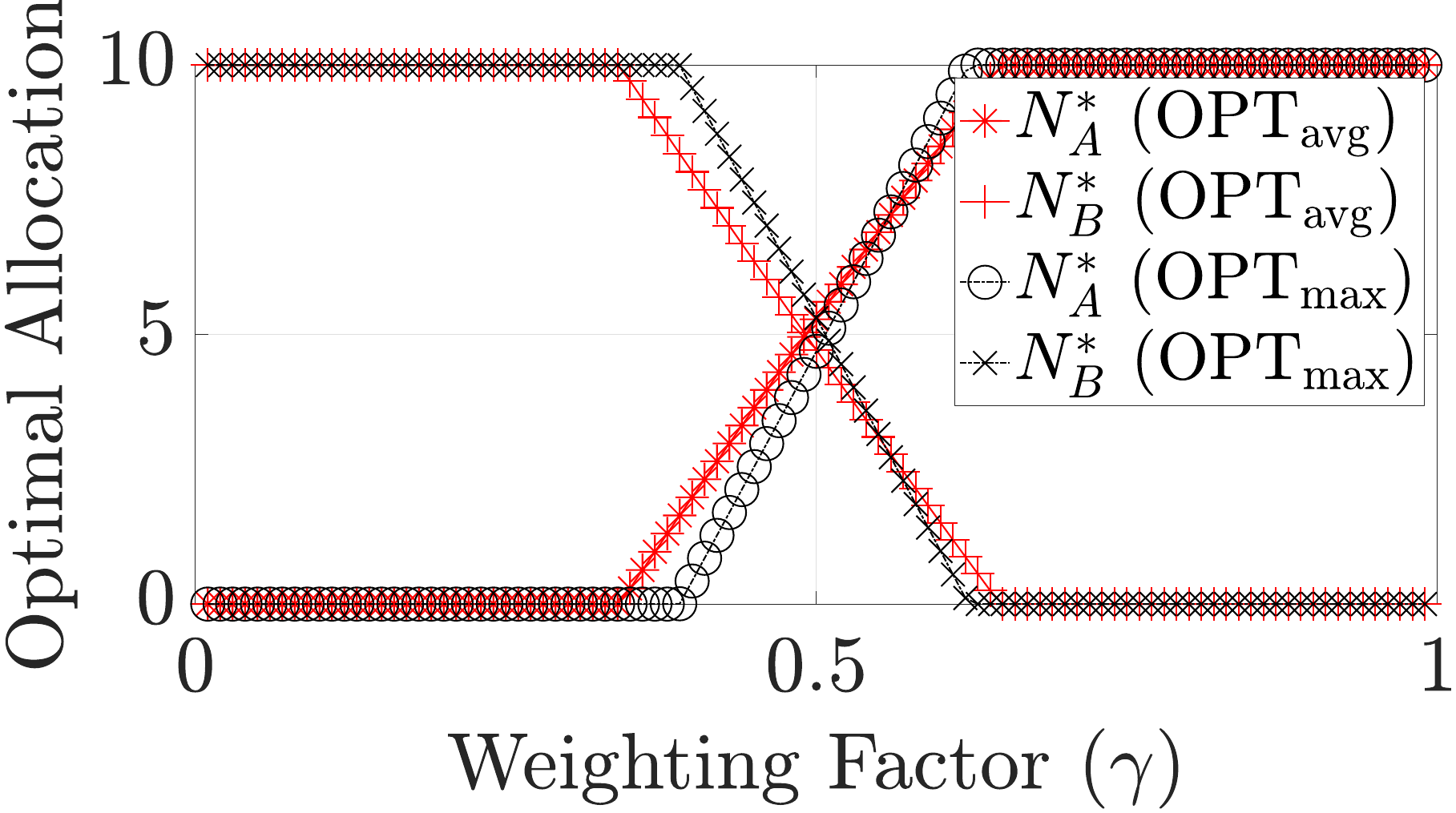}
         \caption{Optimal Allocation}
         \label{fig:Optimal_N_10}
     \end{subfigure}
     \hfill
     \begin{subfigure}[b]{0.31\textwidth}
         \centering
         \includegraphics[width=\textwidth]{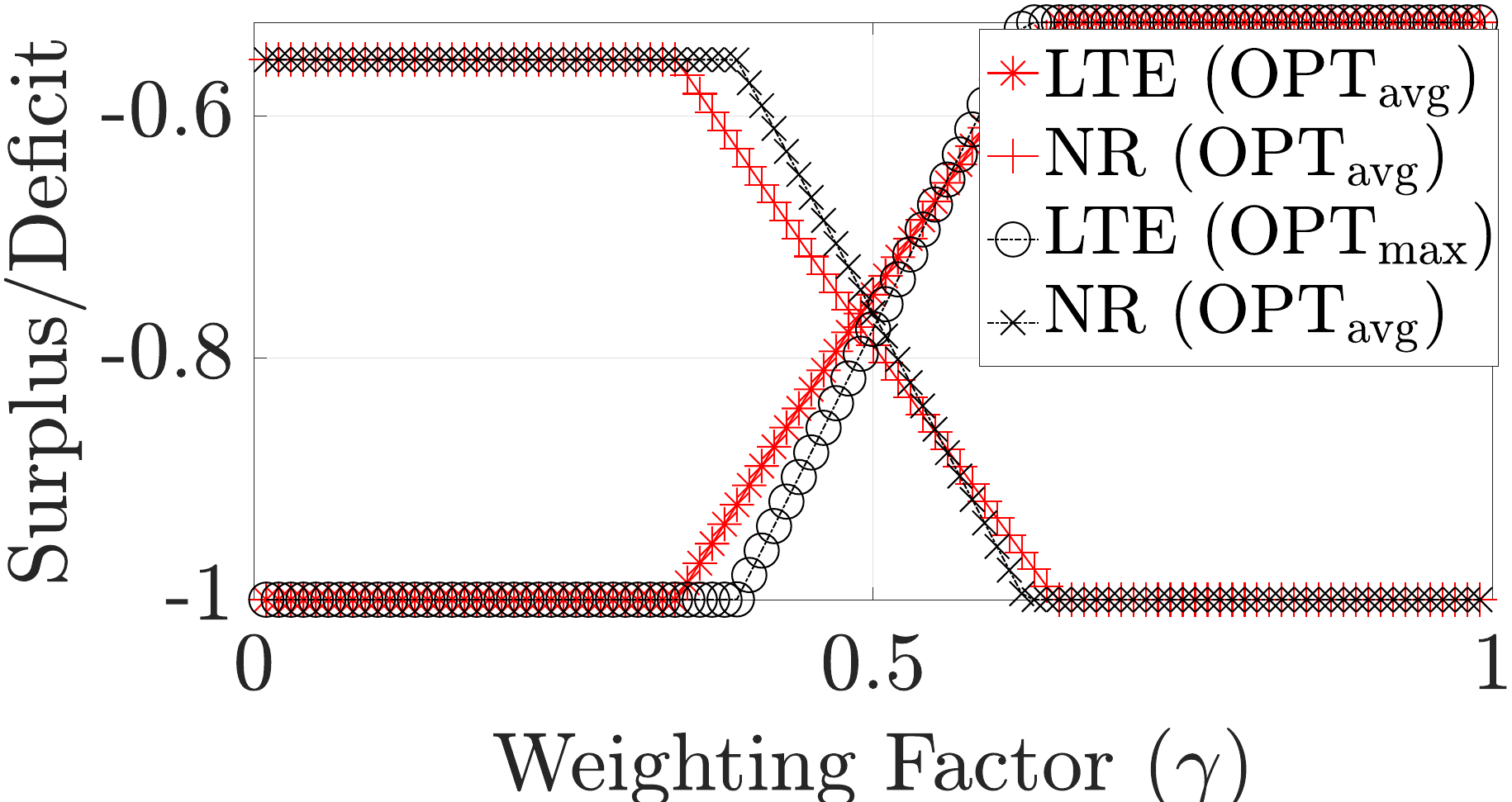}
         \caption{Surplus/Deficit}
         \label{fig:Surplus_Deficit_N_10}
     \end{subfigure}
     \hfill
     \begin{subfigure}[b]{0.31\textwidth}
         \centering
         \includegraphics[width=\textwidth]{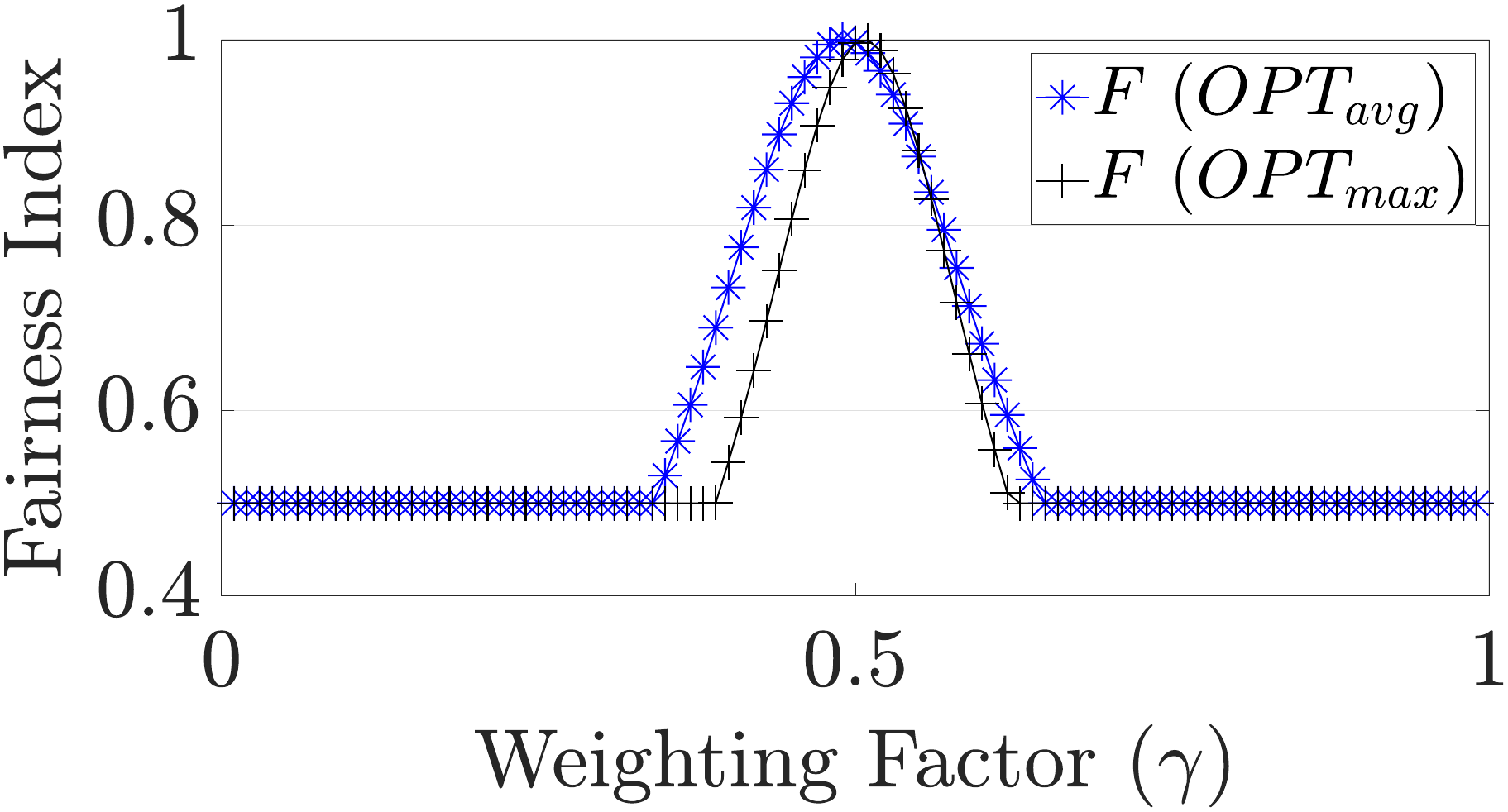}
         \caption{Jain's Fairness Index}
         \label{fig:Jain_Fairness_N_10}
     \end{subfigure}
    \vspace{-0.5em}
    \caption{\small Results for $\OPTP$ versus $\OPTA$ for resource pool size $N_r = 10$~resources.}
        \label{fig:results}
    \vspace{-1.5em}
\end{figure*}

Next, we assess the effectiveness of our resource allocation scheme. We conducted a comprehensive analysis encompassing three distinct resource pool sizes, $N_r \in \{10, 40, 50\}$ corresponding to limited, moderate, and adequate resource availability. We varied the parameter $\gamma$ within each pool size configuration between $0$ and $1$. For each value of $\gamma$, we leveraged statistical insights derived from the predicted demand profiles of $\ranA$ and $\ranB$ to determine the optimal partition sizes, $N_A^*$ and $N_B^*$, under two distinct allocation strategies: $\OPTA$ (utilizing demand statistics like means and variances) and $\OPTP$ (relying on demand maxima). Our evaluation aimed to capture the average surplus or deficit (see Eq.~\eqref{eq:surplus_deficit}) under each scenario, as well as Jain's fairness index (see Eq.~\eqref{eq:JFI}), for different $\gamma$ values and across various resource pool sizes ($N_r$). In the interest of brevity, we present results corresponding to the predicted data for \gls{lte} and \gls{nr} obtained through \gls{arima} and \gls{convlstm}, respectively, at a granularity of 1~hour when $N_r = 10$~resources, as depicted in Fig.~\ref{fig:results}. It is important to note that our insights 
hold consistently across different granularities. Key statistics characterizing the demand profiles of $\ranA$ and $\ranB$ include the following metrics: the mean demand, $\mu_{D_A} = 21.52$ and $\mu_{D_B} = 22.80$; the variances, $\sigma_{D_A}^2 = 12.37$ and $\sigma_{D_B}^2 = 8.33$; and the maxima, $M_{D_A} = 26.31$ and $M_{D_B} = 25.80$. Consequently, the expected total demand approximates 45~resources per hour, while the maximum total demand approaches 53~resources per hour. 

In scenarios with $N_r=10$~resources, where resources are scarce and lag behind demand, both networks face starvation. Total starvation, which occurs when the networks receive no resources, corresponds to a fractional deficit of~$-1$. Our findings (Fig.~\ref{fig:Surplus_Deficit_N_10}) indicate that $\OPTA$ mitigates starvation over a wider range of $\gamma$ values compared to $\OPTP$. Furthermore, $\OPTA$ achieves fairer allocation across a broader spectrum of $\gamma$ values (Fig.~\ref{fig:Jain_Fairness_N_10}). For $N_r=40$~resources, with a moderate pool size that can accommodate mean and maximum demands but falls short of the aggregate demand, $\OPTA$ achieves a perfectly fair allocation ($F=1$) within a specific range of $\gamma$ values ($\gamma \in [0.76, 0.87]$), surpassing the fairness achieved by $\OPTP$ ($\gamma \in [0.47, 0.49]$). However, $\OPTA$ may result in 
deficits across all $\gamma$ values, while $\OPTP$ may result in surpluses, especially when one network has significantly higher priority (near $\gamma$ values of 0 or 1), making $\OPTP$ advantageous in such cases. Finally, for $N_r=50$~resources, where the pool comfortably accommodates both mean and maximum demands, $\OPTP$ outperforms $\OPTA$ for both networks, leading to surpluses. As the resource pool size approaches the maximum total demand, $\OPTA$ tends to allocate resources equal to the mean demand, resulting in deficits, whereas $\OPTP$ results in over-provisioning.

In essence, our results underline the significance of $\gamma$ in resource allocation decisions and demonstrate that the choice between $\OPTA$ and $\OPTP$ depends on the resource pool size relative to network demands. $\OPTA$ excels in fairness but may cause deficits in limited resource scenarios. Conversely, $\OPTP$ is preferable when resources are more abundant as it yields surpluses. These findings provide valuable insights to the controller, enabling it to fine-tune resource allocation strategies to suit diverse network conditions, thereby enhancing overall performance and user satisfaction.

%% file: Conclusion.tex
\section{Conclusions}
\label{sec:conclusion}
In this work, we introduced \gls{prosas}, an \gls{oran}-compatible solution for intelligent spectrum sharing between \gls{lte} and \gls{nr} networks. \gls{prosas} focuses on proactive, intent-driven spectrum management while minimizing resource surplus or deficit experienced by both networks. We demonstrated its effectiveness using real-world \gls{lte} resource usage data and synthetically generated \gls{nr} data. Using these datasets, we developed an optimization scheme and outlined a deployment scenario aligning with \gls{oran} requirements. Our emphasis on real-world data and alignment with industry standards makes \gls{prosas} both practical and pertinent for network operators.

%% file: Appendix.tex
\subsection{Proof of Convexity for $\OPTP$}
\label{app:proof1} 
In this section, we show that $\OPTP$ is a convex-optimization problem and hence has a global minimum.
Mathematically, $\OPTP$ is defined as:
{\small
\begin{align}
\footnotesize
\OPTP:&\min_{N_{A}, N_{B}} J_{max}(N_{A}, N_{B} | \gamma, M_{D_{A}}, M_{D_{B}}),\nonumber\\
\textbf{s.t.}&\quad N_{A} + N_{B} \leq N_{R},\nonumber\\
&\quad 0 \leq N_{A} \leq N_{R}, \nonumber\\
&\quad 0 \leq N_{B} \leq N_{R},
\label{eq:opt_max}
\end{align}}
where, $M_{D_{A}}$ and $M_{D_{B}}$ denote the maximum demand observed by the controller for each network and the objective function $J_{max}$ is defined as:
{\small
\begin{align*}
\small
J_{max} = \gamma\left(\frac{N_A-M_{D_A}}{M_{D_A}}\right)^2 
+ (1-\gamma) \left(\frac{N_B-M_{D_B}}{M_{D_B}}\right)^2.
\end{align*}}
To show that (\ref{eq:opt_max}) is a convex-optimization problem, we first compute the Hessian of its objective function and show that it is positive semidefinite. 
The Hessian of the objective function $J_{max}$ is:
\begin{align*}
H = 
\begin{bmatrix}
\dfrac{\partial^2 J_{max}}{\partial N_A^2} &  \dfrac{\partial^2 J_{max}}{\partial N_A \partial N_B} \cr  \dfrac{\partial^2 J_{max}}{\partial N_B \partial N_A} & \dfrac{\partial^2 J_{max}}{\partial N_B^2}
\end{bmatrix} = 
\begin{bmatrix}
    \dfrac{2\gamma}{{M_{D_A}}^2} & 0 \cr 0 & \dfrac{2(1-\gamma)}{{M_{D_B}}^2}
\end{bmatrix}.
\end{align*}
For a Hessian of the bivariate function $F(x,y)$ to be positive semidefinite, as per Sylvester's criterion, the following two conditions must be satisfied:
\begin{enumerate}[(a)]
    \item $\dfrac{\partial^2 F}{\partial x^2} \geq 0$, and
    \item $\dfrac{\partial^2 F}{\partial x^2}\dfrac{\partial^2 F}{\partial y^2} - \left(\dfrac{\partial^2 F}{\partial x \partial y}\right)^2 \geq 0$.
\end{enumerate}
Since, both (a) and (b) are satisfied for the objective function $J_{max}$, its Hessian is positive semidefinite, and hence it is convex. Also, the inequality constraints are convex, which makes the optimization problem, $\text{OPT}_{\text{max}}$, a convex optimization problem. 

\subsection{Proof of Convexity for $\OPTA$}
\label{app:proof2}
In this section, we show that $\OPTA$ is a convex-optimization problem and hence has a global minimum.
Mathematically, $\OPTA$ is defined as:
{\small
\begin{align}
\footnotesize
\OPTA:&\min_{N_{A}, N_{B}} J_{max}(N_{A}, N_{B} | \gamma, \mu_{D_{A}},\sigma_{D_{A}}^2, \mu_{D_{B}},\sigma_{D_{B}}^2),\nonumber\\
\textbf{s.t.}&\quad N_{A} + N_{B} \leq N_{R},\nonumber\\
&\quad 0 \leq N_{A} \leq N_{R}, \nonumber\\
&\quad 0 \leq N_{B} \leq N_{R},
\label{eq:opt_avg}
\end{align}}
where, $\mu_{D_{A}}$ and $\mu_{D_{B}}$ correspond to the mean and $\sigma_{D_{A}}^2$ and $\sigma_{D_{B}}^2$ correspond to the variance of $\ranA$ and $\ranB$'s demand, respectively, and the objective function $J_{avg}$ is defined as:
{\small
\begin{align*}
J_{avg} = & E\left[\gamma\left(\frac{N_A-D_A}{D_A}\right)^2 
+ (1-\gamma) \left(\frac{N_B-D_B}{D_B}\right)^2\right].
\end{align*}}
On solving the expectation we get:
{\small
\begin{align*}
J_{avg} = & E\left[\gamma\left(\frac{N_A^2}{D_A^2} + 1 -\frac{2N_A}{D_A}\right)+(1-\gamma)\left(\frac{N_B^2}{D_B^2}+1-\frac{2N_B}{D_B}\right)\right],\\
=&1 + \gamma N_{A} \left(N_{A} E\left[\frac{1}{D_{A}^2}\right] - 2E \left[\frac{1}{D_{A}}\right] \right)\\
&+(1-\gamma) N_{B} \left(N_{B} E\left[\frac{1}{D_{B}^2}\right]-2E\left[\frac{1}{D_{B}}\right]\right).
\end{align*}}
The second order Taylor series approximation of the expectation of $1/D_{A}$, $1/D_{A}^2$, $1/D_{B}$ and $1/D_{B}^2$ are as follows:
{\small
\begin{align*}
E\left[\frac{1}{D_A}\right] & = & \frac{1}{\mu_{D_A}} + \frac{\sigma_{D_{A}}^2}{\mu_{D_A}^3}, \quad E\left[\frac{1}{D_{A}^2}\right]  =  \frac{1}{\mu_{D_A}^2} + \frac{3\sigma_{D_{A}}^2}{\mu_{D_A}^4},\\
E\left[\frac{1}{D_B}\right] & = & \frac{1}{\mu_{D_B}} + \frac{\sigma_{D_{B}}^2}{\mu_{D_B}^3}, \quad E\left[\frac{1}{D_{B}^2}\right]  =  \frac{1}{\mu_{D_B}^2} + \frac{3\sigma_{D_{B}}^2}{\mu_{D_B}^4}.
\end{align*}}
On substituting the above expectation values in the objective function, we get:
{\small
\begin{align*}
J_{avg} &= 1+ \gamma N_{A} \left(N_{A} \left(\frac{1}{\mu_{D_{A}}^2} + \frac{3\sigma_{D_{A}}^2}{\mu_{D_{A}}^4} \right) -2\left(\frac{1}{\mu_{D_{A}}} + \frac{\sigma_{D_{A}}^2}{\mu_{D_{A}}^3} \right) \right)\\
&+(1-\gamma) N_{B} \left(N_{B} \left(\frac{1}{\mu_{D_{B}}^2} + \frac{3\sigma_{D_{B}}^2}{\mu_{D_{B}}^4} \right) -2\left(\frac{1}{\mu_{D_{B}}} + \frac{\sigma_{D_{B}}^2}{\mu_{D_{B}}^3} \right) \right).
\end{align*}}
To show that (\ref{eq:opt_avg}) is a convex optimization problem and hence has a global minimum, we first compute the Hessian of the objective function and show that it is positive semidefinite.
The Hessian of the objective function, $J_{avg}$ is:
{\small
\begin{align*}
H 
&= \begin{bmatrix}
    2\gamma\left(\dfrac{1}{\mu_{D_A}^2} + \dfrac{3\sigma_{D_A}^2}{\mu_{D_A}^4}\right) & 0 \cr 0 & 2(1-\gamma)\left(\dfrac{1}{\mu_{D_B}^2}+\dfrac{3\sigma_{D_B}^2}{\mu_{D_B}^4}\right)
\end{bmatrix}.
\end{align*}
}
Since both conditions of Sylvester's criterion are met for the objective function $J_{avg}$, its Hessian is positive semidefinite. Therefore, it is a convex function. Additionally, the inequality constraints are also convex, making the optimization problem, $\OPTA$, a convex optimization problem.